\documentclass[twocolumn]{aastex62}
\usepackage{amsmath,amstext}

\newcommand{\obj}{2007 TC$_{434}$, 2015 KE$_{172}$}

\shorttitle{OSSOS IX: two objects in Neptune's 9:1 resonance}

\begin{document}

\title{OSSOS IX: two objects in Neptune's 9:1 resonance -- implications for resonance sticking in the scattering population} 

\author[0000-0001-8736-236X]{Kathryn Volk}
\correspondingauthor{Kathryn Volk}
\email{kvolk@lpl.arizona.edu}
\affiliation{Lunar and Planetary Laboratory, University of Arizona, 1629 E University Blvd, Tucson, AZ 85721, USA}

\author[0000-0001-5061-0462]{Ruth A. Murray-Clay}
\affiliation{Department of Astronomy and Astrophysics, University of California, Santa Cruz, 1156 High St, Santa Cruz, CA 95064, USA}

\author{Brett J. Gladman}
\affiliation{Department of Physics and Astronomy, University of British Columbia, Vancouver, BC V6T 1Z1, Canada}

\author[0000-0001-5368-386X]{Samantha M. Lawler}
\affiliation{NRC-Herzberg Astronomy and Astrophysics, National Research Council of Canada, 5071 West Saanich Rd, Victoria, British Columbia V9E 2E7, Canada}

\author[0000-0002-5228-7176]{Tze Yeung Mathew Yu}
\affiliation{Department of Physics and Astronomy, University of California, Los Angeles, CA, USA}

\author[0000-0003-4143-8589]{Mike Alexandersen} 
\affiliation{Institute of Astronomy and Astrophysics, Academia Sinica; 11F of AS/NTU Astronomy-Mathematics Building, Nr. 1 Roosevelt Rd., Sec. 4, Taipei 10617, Taiwan, R.O.C.}

\author[0000-0003-3257-4490]{Michele T. Bannister}
\affiliation{Astrophysics Research Centre, Queen's University Belfast, Belfast BT7 1NN, United Kingdom}

\author[0000-0001-7244-6069]{Ying-Tung Chen}
\affiliation{Institute of Astronomy and Astrophysics, Academia Sinica; 11F of AS/NTU Astronomy-Mathematics Building, Nr. 1 Roosevelt Rd., Sec. 4, Taipei 10617, Taiwan, R.O.C.}

\author[0000-0001-9677-1296]{Rebekah I. Dawson}
\affiliation{Department of Astronomy \& Astrophysics, Center for Exoplanets and Habitable Worlds, The Pennsylvania State University, University Park, PA 16802, USA}

\author{Sarah Greenstreet}
\affiliation{Las Cumbres Observatory, 6740 Cortona Dr., Suite 102, Goleta, CA 93117, USA}
\affiliation{Department of Physics, University of California, Santa Barbara, Broida Hall, Santa Barbara, CA, USA}

\author{Stephen D. J. Gwyn}
\affiliation{NRC-Herzberg Astronomy and Astrophysics, National Research Council of Canada, 5071 West Saanich Rd, Victoria, British Columbia V9E 2E7, Canada}

\author[0000-0001-7032-5255]{J. J. Kavelaars}
\affiliation{NRC-Herzberg Astronomy and Astrophysics, National Research Council of Canada, 5071 West Saanich Rd, Victoria, British Columbia V9E 2E7, Canada}
\affiliation{Department of Physics and Astronomy, University of Victoria, Elliott Building, 3800 Finnerty Rd, Victoria, BC V8P 5C2, Canada}

\author[0000-0001-7737-6784]{Hsing~Wen~Lin}
\affiliation{Institute of Astronomy, National Central University, Taoyuan 32001, Taiwan}
\affiliation{Department of Physics, University of Michigan, Ann Arbor, MI 48109, USA}

\author[0000-0003-0926-2448]{Patryk Sofia Lykawka}
\affiliation{School of Interdisciplinary Social and Human Sciences, Kindai University, Japan}

\author[0000-0003-0407-2266]{Jean-Marc Petit}
\affiliation{Institut UTINAM UMR6213, CNRS, Univ. Bourgogne Franche-Comt\'e, OSU Theta F25000 Besan\c{c}on, France}

\begin{abstract}
We discuss the detection in the Outer Solar System Origins Survey (OSSOS) of two objects in Neptune's distant 9:1 mean motion resonance at semimajor axis $a\approx~130$~au. 
Both objects are securely resonant on 10~Myr timescales, with one securely in the 9:1 resonance's leading asymmetric libration island and the other in either the symmetric or trailing asymmetric island.
These objects are the largest semimajor axis objects with secure resonant classifications, and their detection in a carefully characterized survey allows for the first robust resonance population estimate beyond 100~au.
The detection of these objects implies a 9:1 resonance population of $1.1\times10^4$ objects with $H_r<8.66$ ($D~\gtrsim~100$~km) on similar orbits (95\% confidence range of $\sim0.4-3\times10^4$). 
Integrations over 4~Gyr of an ensemble of clones spanning these objects' orbit fit uncertainties reveal that they both have median resonance occupation timescales of $\sim1$~Gyr.
These timescales are consistent with the hypothesis that these objects originate in the scattering population but became transiently stuck to Neptune's 9:1 resonance within the last $\sim1$~Gyr of solar system evolution. 
Based on simulations of a model of the current scattering population, we estimate the expected resonance sticking population in the 9:1 resonance to be 1000-4500 objects with $H_r<8.66$; 
this is marginally consistent with the OSSOS 9:1 population estimate.
We conclude that resonance sticking is a plausible explanation for the observed 9:1 population, but we also discuss the possibility of a primordial 9:1 population, which would have interesting implications for the Kuiper belt's dynamical history.
\vspace{40pt}

\end{abstract}


\section{Introduction}
\label{s:intro}

Observations indicate that our solar system hosts a substantial population of objects that orbit the Sun with large semimajor axes ($a>50$~au) while trapped in mean motion resonances with Neptune \cite[e.g.,][]{Chiang:2003,Lykawka:2007c,Gladman:2008,Gladman:2012,Bannister:201692, Holman:2017}.  
These bodies may have been emplaced by either primordial processes or by current dynamics affecting the large-$a$ objects, with differing implications for the  orbital characteristics and size of each resonance's population. 
Here, we discuss the first two detections of trans-Neptunian objects (TNOs) in Neptune's 9:1 mean motion resonance.  
Both of these objects were discovered by the Outer Solar System Origins Survey (OSSOS; \citealt{Bannister:2016cp,Bannister:2018}) and have precisely determined orbits (see Table~\ref{tab:orbits} for orbital parameters). 
These objects have the most distant orbits of any securely resonant objects known to date, and their detection in a survey with well-characterized detection biases allows for the first robust population estimate of a resonance beyond 100~au.  
We test whether these observed TNOs are consistent with the simplest explanation for resonance occupation at large semimajor axis: transient dynamical sticking.

Objects in Neptune's 9:1 resonance orbit the Sun with semimajor axes of $a \approx a_N(9/1)^{2/3}\approx 130$~au, where $a_N\approx30$~au is the semimajor axis of Neptune's orbit.  
Our two discoveries thus sample a distant population of TNOs.
However, their large orbital eccentricities ($e\approx0.66$ and $e\approx0.7$) bring their perihelia to $q\approx39$ and $q\approx44$~au; this both enables their detection in a flux-limited survey (both objects are currently near perihelion where they are brightest) and allows the possibility of an origin in the closer-in Kuiper belt before being scattered out to their current larger semimajor axes.

Occupation of distant mean motion resonances, particularly $n$:1 resonances, is not surprising in itself.  
TNOs with perihelia interior to $\sim$38~au experience significant orbital perturbations due to encounters with Neptune \citep[e.g.,][]{Duncan:1995,Gladman:2002Icar}.
These perturbations, which may be individually small, cause a random walk in energy and angular momentum \citep{DQT:1987}, which on average increases both the semimajor axis and the eccentricity of the TNO orbits while keeping perihelion distances roughly fixed. 
TNOs with even more distant perihelia ($q\sim40-50~au$) can also exhibit diffusion in their orbits due to extremely weak perturbations at perihelion \citep[see, e.g.,][]{Bannister:2017l91}.
Objects in this ``scattering'' population that happen to achieve $a>$1000 au can then have their perihelia raised by galactic tides and join the Oort cloud \citep[see, e.g.,][]{Dones:2004,Gladman:2005Sci...307...71G}.
Along the way, scattering objects spend a fraction of their time transiently ``stuck'' in mean motion resonances with Neptune for timescales of thousands to millions of orbits \citep[first noted by][]{Duncan:1997}. 
Resonance sticking of Centaurs has been seen in numerical simulations of observed Centaurs \citep[e.g.,][]{Bailey:2009}.
 In particular, several percent of the overall Centaur population is expected to be stuck to the 1:1 co-orbital resonances of Neptune and Uranus at any given time \citep{Alexandersen:2013}.
Transient sticking of large-$a$ TNOs is particularly efficient in Neptune's $n$:1 resonances \citep{Gallardo:2006,Lykawka:2007Icar}, and at large orbital separations the total transiently-stuck population is expected to be comparable to the population of actively scattering objects \citep{Yu:2018}. 
In numerical explorations of transient sticking in the scattering population, \citet{Lykawka:2006} and \cite{Lykawka:2007Icar} reported finding particles sticking in the 9:1 resonance.

We demonstrate in Section~\ref{s:orbint} that both 9:1 objects discovered by OSSOS occupy resonant orbits that are stable on timescales approaching or exceeding 1~Gyr.
Such long timescale resonance occupation is not unexpected based on resonance sticking simulations; \citet{Lykawka:2006} found one of their 9:1 sticks to last 900 Myr.
Though most individual resonance sticks last for only a short time (a few libration periods),
\citet{Yu:2018} find that in a time-averaged population, there is a roughly equal likelihood of seeing a sticking event per log bin in stick timescale; i.e., long timescale sticks are relatively rare compared to short timescale ones, but their long-lived nature and the fact that the scattering population is decaying over time (and thus more objects were available in the past to stick to resonances) means that short and long timescales are roughly equally likely to be observed today.

In this paper, we take advantage of the well-characterized nature of OSSOS to examine whether our newly discovered 9:1 objects are consistent with models of transient sticking.  
As described in \cite{Bannister:2016cp}, OSSOS is designed to allow robust comparisons between observed and simulated populations of TNOs.  
We discuss the orbits of the two OSSOS 9:1 objects in Section~\ref{s:orbint}, describing their short- and long-term dynamics in the resonance.
We then use the OSSOS survey simulator to estimate the size of the intrinsic 9:1 population based on our two detections (Section~\ref{s:survsim}). 
In Section~\ref{s:origin}, we discuss the possible origins of this 9:1 population. 
We estimate the expected number of transiently stuck 9:1 objects based on theoretical modeling and observational constraints on the population of current actively scattering objects for comparison to the OSSOS 9:1 population estimate (Section~\ref{s:stickmodel}). 
We discuss the possibility that the 9:1 population was captured very early in the solar system's history, either by transient sticking from the massive primordial scattering population or during the era of planet migration (Section~\ref{s:primordial}).
We also compare the observed 9:1 population to population estimates of other distant resonant populations (Section~\ref{s:rescomparison}).
Section~\ref{s:sum} provides a summary of our findings and the implications these two detections have for resonance sticking in the scattering population.

\begin{deluxetable*}{cclllrrllcc}[h]
\tabletypesize{\scriptsize}
\tablecaption{Barycentric orbit fit and short-term resonant dynamics for the OSSOS 9:1 objects \label{tab:orbits}}
\tablehead{
\colhead{OSSOS} & \colhead{MPC} & \colhead{$a$} & \colhead{$e$} & \colhead{$q$} & \colhead{$i$} & \colhead{$\phi$ center} & \colhead{$A_{\phi}$} &  \colhead{$H_r$} & \colhead{observed} & \colhead{number of} \\[-8pt] 
\colhead{designation} & \colhead{designation\tablenotemark{a}} & \colhead{au} & \colhead{}& \colhead{au} & \colhead{deg.} & \colhead{deg.} & \colhead{deg.} &  \colhead{} & \colhead{arc (yr)} & \colhead{observations}
}
\startdata
o5m72  & 2015 KE$_{172}$ & $129.80 \pm 0.026$ 		& 0.6600 	& 44.13 & 38.361 & $\sim260$ & $\sim\phm{1}60$ &  8.20 & 3.26 & 36\\ 
  &  & 	& 	& &  & \phm{$\sim$}180  & $\sim160$ & \\ 
  &  & 	& 	& &  &      &           & \\    
o4h39 & 2007 TC$_{434}$ & $129.93 \pm 0.03$ 		& 0.6952 	& 39.60 & 26.468 & $81 \pm 3$ & $35^{+6}_{-7}$ &  7.13 & 4.06 & 42\tablenotemark{b}\\ 
\\
\enddata
\tablecomments{Best-fit barycentric orbital elements were determined using the \citet{Bernstein:2000} orbit fitting code and all available OSSOS astrometry through October 2017.  We only explicitly list the uncertainty in the semimajor axis for these orbit fits. For the eccentricity and inclination, all the digits in the table are significant, so the (very small) uncertainties are in the next digit. The libration centers and libration amplitudes were determined from a 10~Myr integration of 250 clones spanning the uncertainties of the orbit fit.  Based on this analysis, o5m72 could be in either the symmetric or trailing asymmetric libration island, so approximate libration amplitudes are given for each of those islands. We do not list uncertainties for these parameters because the libration is not well-enough behaved to easily define them; the reader is instead referred to Figure~\ref{f:current-dynamics}, which shows the distribution of these parameters.  The uncertainties in o4h39's libration center and amplitude are 1-sigma uncertainties taken from the distribution of the 250 clones. 
}
\tablenotetext{a}{
The Minor Planet Center Electronic Circulars for these two objects are \citet{o4h39MPEC} and \citet{o5m72MPEC}. 
}
\tablenotetext{b}{
This number only reflects the number of OSSOS observations. As indicated by the 2007 designation, upon submission to the Minor Planet Center (after the initial submission of this paper), o4h39 was linked to a previously detected object with a 2 day arc and then to additional, previously unpublished observations from 2004. These 15 additional observations are not included in the orbit fit and uncertainty listed above. We briefly discuss the effects of these additional observations on our analysis in Section~\ref{ss:longterm}. 
}
\label{t:orbits}
\end{deluxetable*}

\section{Dynamical Characterization of the observed 9:1 objects}\label{s:orbint}

OSSOS discovered two TNOs (o5m72 and o4h39) that are securely classified 
as librating in Neptune's 9:1 mean motion resonance on 10 Myr timescales (see \citealt{Gladman:2008} for a full description of the classification process and the definition of secure classifications).
All observations of these objects are reported in the full OSSOS data release, along with the complete survey's sensitivity to moving objects as a function of magnitude and rate of motion \citep{Bannister:2018}. Here we discuss the orbits of these objects as well as their current and long-term dynamical evolution.

\subsection{Current Orbit and Resonant Dynamics}\label{ss:shortterm}

Table~\ref{t:orbits} lists the best-fit orbits for o5m72 and o4h39 based on all available OSSOS astrometry through October 2017 along with the libration center and amplitude for the resonant angle $\phi_{9:1} = 9\lambda_{TNO} - \lambda_{Neptune} - 8\varpi_{TNO}$ (where $\lambda$ is the mean longitude and $\varpi$ is the longitude of perihelion); the libration amplitude is defined as $A_{\phi} = (\phi_{9:1, max}-\phi_{9:1, min})/2$. 
We note that, following typical dynamical conventions \citep[such as those in, e.g.,][]{MurrayDermott:1999}, this resonant argument would be described as being $8^{th}$ order in the TNO's eccentricity;
 in this theoretical context, one might expect the phase space volume of this resonance to be too small to host a significant number of resonant objects.
Numerical simulations of transient resonance sticking show, however, that distant $n$:1 resonances in fact host the largest populations of transiently-stuck objects \citep{Lykawka:2007Icar,Yu:2018}. This apparent disconnect is due to the fact that our typical interpretation of resonance order in the solar system is based upon the assumption of small eccentricities. In the small eccentricity limit, the order of the resonance describes the number of planet-TNO conjunctions that occur over the course of a single resonance cycle; high-order resonances lead to large numbers of conjunctions, weakening the resonant perturbation. However, in the case of distant, highly eccentric TNO orbits, the vast majority of conjunctions occur at such large physical separations between the planet and the TNO that they are dynamically unimportant; the most important interaction between the TNO and Neptune occurs when the TNO is at perihelion.
TNOs in $n$:1 resonances have one perihelion passage per resonance cycle, $n$:2 resonant TNOs have two, and so on. This has led to the suggestion 
\citep[e.g.,][]{Pan:2004} that distant $n$:1 resonances should be called `first' order resonances rather than $(n-1)$ order resonances. 
Furthermore, among resonances at similar orbital separations, average transient sticking timescales are largest for $n$:1 resonances, followed by $n$:2, and so on.  \citet{Yu:2018} suggest that these longer stick times result from longer resonance libration periods.  
Since simulations predict that the instantaneous population of transiently-resonant scattering objects is largest for $n$:1 resonances, the detection of objects in the 9:1 is not surprising.   
The objects described here are the most distant of these $n$:1 objects yet to be securely identified via an analysis of the orbital element uncertainty, but we expect future observations to reveal many more resonant objects in even more distant resonances.

The best-fit orbit and the uncertainties in the orbital parameters for o4h39 and o5m72  are taken from the \citet{Bernstein:2000} orbit fitting code. The libration characteristics are determined from a 10~Myr integration of the best-fit orbit and 250 clones whose orbits span the uncertainty range for the best-fit orbit; these additional clones are generated using the covariance matrix for the orbital parameter uncertainties generated by the \citet{Bernstein:2000} orbit fitting code.  All integrations were performed using the rmvs3 subroutine in the SWIFT numerical integration package \citep{Levison1994}. The integrations included the four giant planets and the Sun (augmented by the mass of the terrestrial planets) as the only massive bodies in the system, and the planets initial positions and velocities were taken from JPL Horizons \citep{jplhorizons} for the epoch for the orbit fits. All clones of o5m72 and o4h39 were included as massless test particles in the simulations. 
We used a base integration step size of 0.5 years (with smaller, adaptive time-steps used if planets are approached), 
and test particles were discarded if they reached heliocentric distances larger than 1500~au or smaller than 5~au. Orbital elements were output every 1000 years for analysis of the resonant behavior.

\begin{figure*}[htbp]
\includegraphics[width=7.0in]{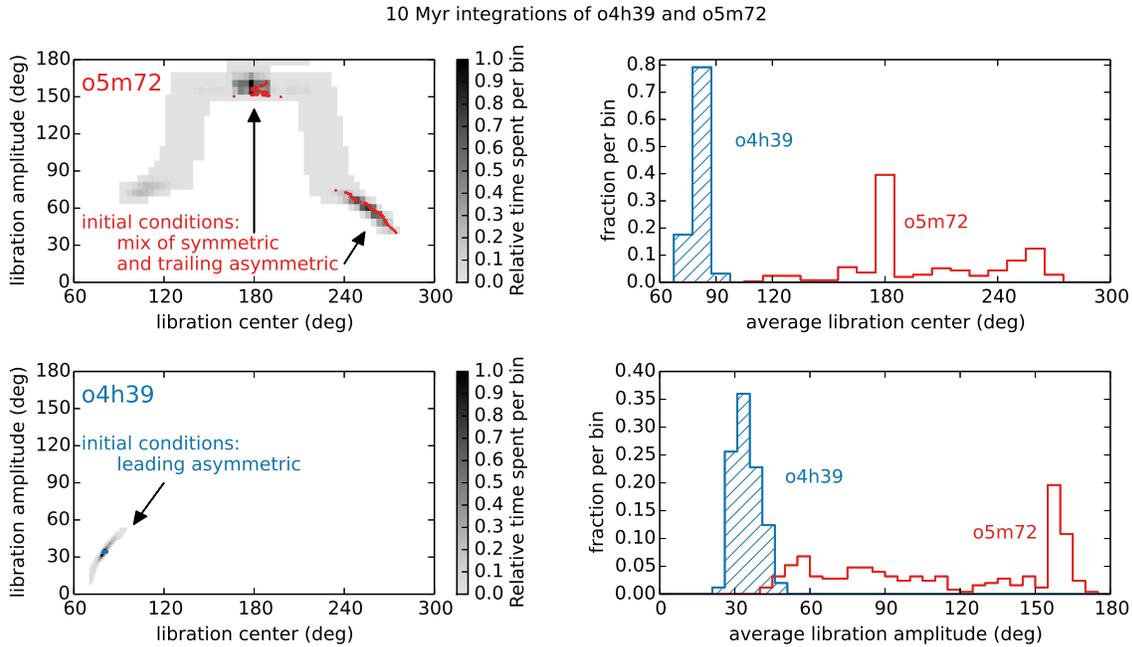}
\caption{Left: Time weighted distribution of libration amplitude vs. libration center for 250 clones of o5m72 (top) and o4h39 (bottom) over a 10 Myr integration (in the gray-scale map, white indicates unexplored phase space and black is where the clones spend the most time); the colored dots show the initial distribution of each objects' clones (which is representative of the orbit-fit uncertainties) for comparison to the time-averaged distribution. Right: time-averaged histograms of the libration center (top) and libration amplitude (bottom) for the same clones of o5m72 (empty red histogram) and o4h39 (hatched blue histogram). All of the o4h39 clones are small amplitude librators in the 9:1's leading asymmetric island. The clones of o5m72 spend time in all three libration islands, but predominantly switch between symmetric and trailing asymmetric libration.}
\label{f:current-dynamics}
\end{figure*}

Figure~\ref{f:current-dynamics} shows the range of resonant behavior seen in the 10~Myr integrations of both o4h39 and o5m72. 
Like other $n$:1 resonances, the 9:1 has three possible centers of libration for the resonance angle \citep[see., e.g.,][]{Beauge:1994}.
The symmetric librators have an averaged value of $\left<\phi_{9:1}\right> = 180^{\circ}$, which means that over a libration cycle they explore a 
large range of perihelia offsets (relative to Neptune), completely covering the range $180^{\circ}\pm A_{\phi}$ away from Neptune; the libration amplitude for the symmetric libration mode is typically large ($A_{\phi}\gtrsim 100^{\circ}$). 
The two asymmetric libration islands are centered near $\phi_{9:1}\sim90^{\circ}$ (leading asymmetric) and $\phi_{9:1}\sim270^{\circ}$ (trailing asymmetric). 
Asymmetric librators have much smaller libration amplitudes (typically $\lesssim80^\circ$) than symmetric ones. 
The libration period for $\phi_{9:1}$ ranges from $\sim3\times10^4-10^5$~years for these two objects. 
We note that these two objects can experience libration for additional 9:1 resonant arguments beyond the eccentricity-type argument we call $\phi_{9:1}$. For mixed eccentricity-inclination type resonant arguments (e.g., $\phi = 9\lambda_{TNO} - \lambda_{Neptune} - 6\varpi_{TNO} - 2\Omega_{TNO}$), we typically see libration similar to those of $\phi_{9:1}$, but about a libration center that slowly circulates on 50-100~Myr timescales, corresponding to the secular regression of the objects' longitudes of ascending node. Because the eccentricity-type resonance is the strongest and most clearly librating for these highly eccentric TNOs, we only consider the libration amplitudes for $\phi_{9:1}$.

Over the 10 Myr simulations, all of the o4h39 clones are small-amplitude leading asymmetric librators with libration amplitude $A_{\phi} = 35^{+6}_{-7}$ degrees; the uncertainty in $A_{\phi}$ represents the 1-$\sigma$ range from the distribution in the lower right panel of Figure~\ref{f:current-dynamics}. 
The clones of o5m72 are more mobile and spend most of the 10 Myr simulation switching between trailing asymmetric and symmetric libration; some clones also spend time in the leading asymmetric island. 
Thus o5m72 has a less well-defined libration amplitude, and we provide only the median amplitude for the symmetric and trailing asymmetric clones in Table~\ref{t:orbits} rather than an uncertainty. 
Neither object is currently experiencing so-called Kozai libration (libration of the argument of perihelion, $\omega$) within the 9:1 resonance; because this libration occurs on longer timescales than mean motion resonant libration, a 100 Myr integration was used to check for $\omega$ libration.

\subsection{Long-term Dynamics}\label{ss:longterm}

\begin{figure}
\vspace{40pt}
\includegraphics[width=3.2in]{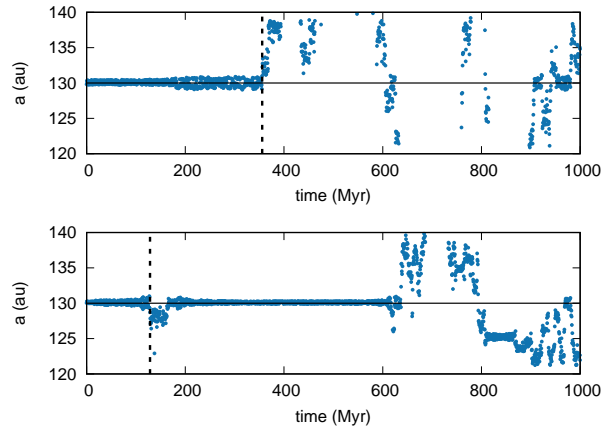}
\caption{Example semimajor axis evolution of two different clones of o4h39. The center of the 9:1 resonance is indicated by the solid black horizontal line. The vertical dashed lines indicate the time at which each clone is considered to have stopped being continuously resonant. The evolution in the bottom panel shows that some clones will re-enter the 9:1 after temporarily leaving.}
\label{f:example}
\end{figure}

To determine how long o4h39 and o5m72 are likely to remain in the 9:1 resonance, we integrated their orbits forward in time for 4 Gyr using 1000 clones that span the orbital uncertainties from the best-fit orbit's covariance matrix. We use the same integration method outlined in Section~\ref{ss:shortterm}, but with an output frequency decreased to every $10^4$ years. Figure~\ref{f:example} shows some example semimajor axis histories for clones of o4h39. We tracked the semimajor axis history of each clone over a sliding 5 Myr time window for the duration of the simulation to determine if or when it left the resonance. We consider a test particle to have left the resonance if the average semimajor axis over the 5 Myr window deviated from the expected resonant value (130.06~au) by more than 0.3~au and the minimum or maximum semimajor axis value over the window deviated by more than 1~au; we take the midpoint of the time window to be the time at which the clone left the resonance.  Some clones leave the resonance but then re-enter it after a sequence of gentle scattering event with Neptune (as shown in the bottom panel of Figure~\ref{f:example}); in these instances we consider the clone to have left the resonance as of the first excursion out of resonance. We use the semimajor axes of the test particles to determine resonance membership rather than the evolution of $\phi_{9:1}$ because individual libration cycles are not fully resolved at this reduced output frequency and high-amplitude libration can be difficult to distinguish from circulation. The semimajor axis limits described above were determined by visually examining the semimajor axis and resonant angle evolution of 100 clones of each object and adjusting the limits until they reproduced the visually determined departure times from the resonance.

\begin{figure}
\includegraphics[width=3.2in]{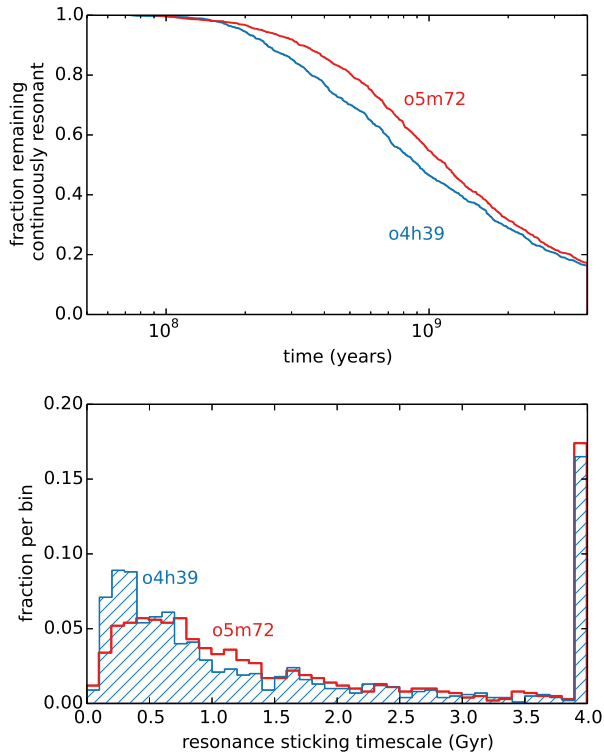}
\caption{Fraction of continuously resonant clones remaining over time (top) and distribution of initial resonance sticking times (bottom) for 1000 clones of o5m72 (red line and empty red histogram) and o4h39 (blue line and hatched blue histogram) in a 4 Gyr simulation.}
\label{f:time}
\end{figure}

Figure~\ref{f:time} shows the distribution of resonance sticking timescales for the clones of o4h39 and o5m72. The clones of o4h39 have a median resonance sticking time of 910 Myr, with $\sim16$\% of the clones remaining continuously resonant at 4 Gyr. An additional $\sim8\%$ of the clones have scattered back into the 9:1 resonance at 4 Gyr. The clones of o5m72 have a median sticking time of 1.37 Gyr, and $\sim17\%$ of the clones are still continuously resonant at 4 Gyr; an additional $\sim6\%$ of the clones have scattered back into the 9:1 at 4 Gyr.

The fact that o4h39 has a slightly shorter sticking timescale despite currently being deeper in resonance than o5m72 is likely at least partly due to its lower perihelion distance ($\sim39$~au) compared to o5m72 ($\sim44$~au). 
However, we note that longer-term variations in eccentricity while o5m72 remains resonant can lower its perihelion to $\sim38$~au, allowing clones to be scattered out of resonance.
Over the 4 Gyr simulations, $\sim31\%$ of the o4h39 clones and $\sim44\%$ of the o5m72 clones were removed from the simulation because they were scattered far enough outward to reach heliocentric distances larger than 1500~au.

Because we infer the sticking timescale from an ensemble of clones, we must consider how the observational uncertainties in each objects' current semimajor axis affects the distribution of likely sticking timescales. (While the eccentricity will also affect stability in the resonance, the uncertainties in $e$ and $a$ are strongly correlated, so we need only examine the variation in sticking timescale across one parameter.) We examined the resonance sticking time as a function of initial semimajor axis for the clones of each object.  For the o4h39 clones, this distribution is very similar across the entire range of $a$ sampled in our simulations. In contrast, the observational uncertainty in o5m72's orbit spans a portion of the 9:1 resonance where the stability timescale is changing; the most stable o5m72 clones are not evenly distributed across o5m72's uncertainty range. Thus, we expect future observations, which will decrease the orbit-fit uncertainties, to have a much more significant effect on the inferred sticking timescale for o5m72 than for o4h39. We can confirm this expectation for o4h39, because (as noted in Table~\ref{t:orbits}) additional observations were linked to o4h39 after the OSSOS observations were sent to the Minor Planet Center after the submission of this paper. These additional observations occurred in 2004 and 2007, and their inclusion in o4h39's best fit orbit approximately halves the semimajor axis uncertainty. The new orbit has short-term evolution in the 9:1 very similar to the orbit fit we used throughout this work, with a libration amplitude well-within our uncertainty range. We performed a 1 Gyr integration of 100 clones of the new orbit and confirm that the fraction of clones remaining in the 9:1 at 1~Gyr is statistically indistinguishable from the fraction shown in Figure~\ref{f:time} for that time.

\section{OSSOS population estimate}\label{s:survsim}

We use the OSSOS survey simulator (described in detail by \citealt{Bannister:2016cp,Bannister:2018,Lawler:2018ss}) to model the detection biases of OSSOS and estimate the intrinsic number of 9:1 objects required to match the two observed ones. 
We estimate the population of 9:1 resonant objects by constructing a model resonant population with eccentricities and inclinations within a small range encompassing the observed objects' values, requiring the survey simulator to generate two detections from this model population, and recording the range of total simulated population sizes required to generate those two detections.  
Specifically, the model population has eccentricities and inclinations distributed uniformly in the ranges $i=$25--40$^\circ$ and $e=0.6$--0.7. 
These uniform ranges are chosen to encompass the observed values given that our two detections cannot strongly constrain the intrinsic 9:1 orbital distribution; 
we also note that because we are not observationally sensitive to very low eccentricity objects in such a distant resonance, we can only model the higher eccentricity population within the resonance.
For the libration islands, we choose a model with half the resonant population librating in the symmetric mode, and then split the asymmetric population equally between the leading and trailing states, similar to the 2:1 resonance model in \citet{Volk:2016};
 this choice is consistent with the two observed objects, clones of which show libration around all three libration centers (see Section~\ref{ss:shortterm}). The locations of the OSSOS pointings on the sky means the survey was observationally sensitive to objects librating around all three 9:1 libration centers, so changing the distribution of libration centers in our model will not strongly affect the population estimate.   
We assume an exponential absolute magnitude distribution $N\propto10^{\alpha H}$ with $\alpha$=0.8 down to $H_r$=8.66 (which corresponds to diameter $D\gtrsim100$~km for typical TNO albedos and colors).
This choice of H magnitude distribution is consistent with observations of the scattering population \citep{Shankman:2016hi,Lawler:2018}.
The simulated detections resulting from this simplified model satisfactorily match the two detections and yield a population estimate of $\sim$11,000 9:1 resonators with $H_r<8.66$, with a factor of 3 uncertainty (at 95\% confidence).

The fact that OSSOS, which is dominated by sky coverage near the ecliptic, finds its two 9:1 resonators to have inclinations of $26^\circ$ and $38^\circ$ might be viewed as surprising.
Ecliptic surveys should be heavily biased towards detection of the lowest orbital inclination objects, which spend much more time in the observed sky area.
If we alter the above model to draw the inclinations from a standard $\sin(i)\times$Gaussian distribution with a width of 20 degrees  \citep{Brown:2001}, then only 20\% of the simulated detections have $i>25^\circ$; 
the resulting 4\% probability that both discovered TNOs are in this range is on the verge of being significant, but is not strong enough (given that the 25$^\circ$ dividing line was determined post-facto from the sample) to justify rejection of this inclination distribution. 
 We note that the \citet{Pike:2015} analysis of the observed 5:1 resonant population came to a similar conclusion about the inclination distribution; they could not reject the $\sin(i)\times$Gaussian distribution, but all of the observed 5:1 objects had inclinations higher than $\sim20^{\circ}$.
For the 9:1 resonance, an off-centered Gaussian (such as found by \citealt{Gulbis:2010}, centered at an inclination of $20^{\circ}$ with a width of $7^{\circ}$) would reduce the tension between the assumed inclination distribution and the observations;  this off-centered Gaussian inclination distribution would drop the nominal population estimate to $\sim7000$ objects with $H_r<8.66$ compared to the uniform inclination range ($i=$25--40$^\circ$) considered above.

\section{Possible origins of the 9:1 objects and comparison to other populations}\label{s:origin}

Here we examine whether the OSSOS 9:1 population estimate is consistent with the hypothesis that the 9:1 population is dominated by objects transiently stuck from the current scattering population (Section~\ref{s:stickmodel}) as well as the possibility of a more primordial origin (Section~\ref{s:primordial}). 
We also compare our 9:1 population estimate to other distant $n$:1 populations (Section~\ref{s:rescomparison}).

\subsection{Transient sticking from the current scattering population}\label{s:stickmodel}

\cite{Lykawka:2006} and \cite{Lykawka:2007Icar} investigated resonance sticking in simulations of actively scattering test particles. 
They identified several captures in the 9:1 with timescales ranging from a few to a hundred Myr, showing both symmetric and asymmetric libration behavior.
These simulations demonstrate that the 9:1 is capable of temporarily capturing scattering objects, but they do not provide sufficient statistics to estimate the current expected population of transient 9:1 objects.

To better estimate the expected transiently stuck 9:1 population, we instead extrapolate the results of \citet{Yu:2018} who investigate resonance sticking in the region $a=30-100$~au. 
The orbital distribution of the initial population in this simulation is based on the \citet{Kaib:2011ky} model of the scattering population, which has been shown to be consistent with observations of the current scattering population \citep{Shankman:2016hi}.
This same model is used to generate the population estimate for the scattering population \citep{Lawler:2018} necessary to ultimately compare the predicted transient 9:1 population to the observationally derived 9:1 population estimate from Section~\ref{s:survsim}.
\citet{Yu:2018} find that starting from this population of actively scattering test particles and looking at the time-averaged population on a timescale of 1 Gyr, $\sim$40\% of 30 $< a <$ 100 au TNOs should be transiently stuck in a resonance at any given time; 
this implies that, at any given time, there are about 1.3 times as many actively scattering (non-resonant) objects as there are objects transiently stuck to resonances in this range.
These simulations indicate that about a quarter of the the transiently stuck particles are in n:1 resonances, and the transient populations of the n:1 resonances increase with semimajor axis.

\begin{figure}[htbp]
\includegraphics[width=3.2in]{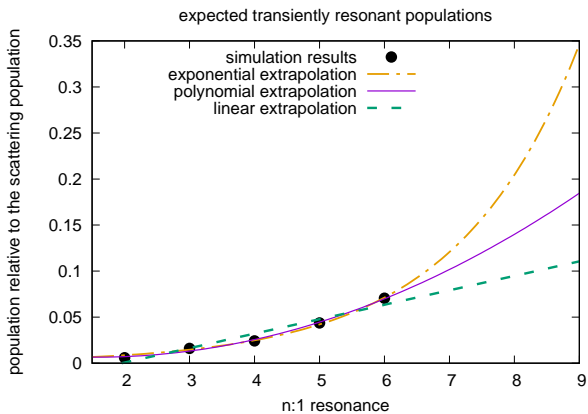}
\caption{Expected transient sticking populations for the $n$:1 resonances in the \citet{Yu:2018} simulations (black dots) relative to the actively scattering population from $30<a<100~au$ . We show three different fits to the simulation data extrapolated out to the 9:1 resonance at $a\simeq$130 au.}
\label{f:extrapolation}
\end{figure}

Figure~\ref{f:extrapolation} shows the number of transiently stuck particles for the n:1 resonances with $a<100$~au in the \citet{Yu:2018} simulations (black dots) in units of the total actively scattering population from $a=30-100$~au. 
We extrapolate these simulation results slightly beyond 100~au to estimate the expected population of transiently stuck 9:1 objects. 
We fit three different functional forms to the  distribution of total sticking populations for the n:1 resonances, $P_{n:1}$: a linear fit ($P_{n:1} = a+ bn$), a second degree polynomial fit ($P_{n:1} = a + bn + cn^2$), and an exponential fit ($P_{n:1} = a\exp{bn}$). 
These fits are shown, extrapolated to $P_{9:1}$, in Figure~\ref{f:extrapolation}. 
The linear fit is poor quality, but we use it as a lower-bound estimate; the polynomial and exponential fits are of similar quality. 
The exponential fit is intended to provide an upper bound, and we 
do not expect any of these extrapolations to hold out to arbitrarily large semimajor axes. 
The ability of Neptune's resonances to efficiently capture scattering objects should drop off for extremely distant resonances.
However, the \citet{Lykawka:2007Icar} simulations show temporary sticking out to $\sim250$~au, including sticks in the 9:1 resonance. 
 Thus we expect our conservatively  wide range of extrapolations out to the 9:1 at $a\sim130$~au to encompass the true 9:1 sticking population.
The extrapolated \citet{Yu:2018} results imply that the population of transiently stuck 9:1 objects is $\sim0.13-0.36$ times as large as the population of actively scattering objects from $a=30-100$~au.

To translate this theoretical relative 9:1 population estimate to an absolute number (to compare with the observationally derived population estimate in Section~\ref{s:survsim}), we need a population estimate for the actively scattering population. 
An analysis of the combined OSSOS and CFEPS actively scattering population detections implies that there are $\sim1.1\pm0.2\times10^4$ scattering objects with $a<100$~au and $H_r<8.66$ \citep{Lawler:2018}. 
This population estimate is model dependent, but uses the same \citet{Kaib:2011ky} orbital distribution for the scattering population as the \citet{Yu:2018} simulations. 
Based on the extrapolation of the \citet{Yu:2018} results, this translates to $\sim1.1-4.5\times10^3$  objects with $H_r<8.66$ transiently stuck in the 9:1 resonance. This should be compared to the $\sim0.4-3.3\times10^4$ (95\% confidence range) population estimate derived from the two OSSOS 9:1 objects in Section~\ref{s:survsim}.
 We note that there are a few potential differences between the resonance sticking population in the scattering simulations and the nominal 9:1 model used to produce our observational population estimate. 
The first is that the inclinations in the \cite{Yu:2018} simulations are generally lower than the $i=$24-40$^\circ$ range used to produce our nominal 9:1 population estimate. 
In the simulations, resonance stickiness does not depend strongly on inclination, so the number of transiently stuck 9:1 objects is unlikely to change with a different inclination distribution; however the observational population estimate does depend on the modeled inclination distribution. 
The comparison between the expected 9:1 sticking population and the observed population improves slightly if we model the 9:1 with an off-center Gaussian inclination distribution, which slightly decreases the observed 9:1 population estimate (see Section~\ref{s:survsim}).
The second possible difference between the simulations and our nominal 9:1 model is the assumed eccentricity/perihelion distance distribution. 
The two observed 9:1 objects currently have perihelion distances that are larger than typical for the actively scattering population that served as the initial conditions for the \citet{Yu:2018} simulations.
However, as noted in Section~\ref{ss:longterm}, eccentricity cycling within the 9:1 causes longer-term variations in $q$ for both objects; during the long-term simulations, clones of both objects do achieve perihelion distances consistent with the scattering population.
Direct simulations of the transient 9:1 population would allow the calculation of a time-averaged perihelion distribution (and thus eccentricity distribution) that could be used in place of our uniform $e=$0.6--0.7 distribution to generate an improved observational population estimate.
However, with only two detections, the uncertainty in the population estimate for any 9:1 model is very large, so we leave this for future investigations. 

We these caveats in mind, we note that the upper range of the theoretically estimated 9:1 transient population ($\sim1.1-4.5\times10^3$ objects with $H_r<8.66$) is consistent with the lower range of the observationally derived population estimate ($\sim0.4-3.3\times10^4$ objects with $H_r<8.66$).
We thus conclude that the two OSSOS 9:1 detections are marginally consistent with the 9:1 population expected to be transiently stuck from the current scattering population.

\subsection{Possible Primordial Origin}\label{s:primordial}

Given the only marginal agreement between the OSSOS 9:1 population estimate and the expected number of 9:1 objects transiently stuck from of the current scattering population, as well as the fact that $\sim15-20\%$ of the o5m72 and o4h39 clones remain resonant on 4 Gyr timescales, it is possible that these objects are the remnant of a larger primordial resonant population. 
The observed 9:1 objects could either have been captured during the era of planet migration or could represent extremely long-timescale ``transient'' sticks from earlier in the solar system's history when the scattering population was much more numerous. 
The primordial scattering population is expected to be one to two orders of magnitude larger than the current scattering population \citep[see, e.g.,][]{Kaib:2011ky,Brasser:2013}. 
The \citet{Yu:2018} simulations only considered 1 Gyr timescales for both population decay and resonance sticking, but we can use these results to estimate whether longer-timescale sticks from the larger primordial population significantly enhance the expected 9:1 population. 

From the \citet{Yu:2018} simulations, we find that the probability of a scattering object (with semimajor axis in the range 30-100~au) getting transiently stuck in a resonance for timescale $t_{stick}$ goes approximately as $P(t_{stick}>t) \propto t^{-1}$ for sticks up to a few hundred Myr in length. The stick probability appears to fall faster for longer sticks ($P(t_{stick}>t) \propto t^{-2}$), although the statistics are poorer.
The decay of the Kuiper belt populations over time is often modeled as $N(t) \propto t^{-b}$; for various models of the scattering population, $b$ values in the range $\sim0.7-1.3$ appear to approximate the population reduction reasonably well \citep[see, e.g.,][]{Kaib:2011ky,Brasser:2013,Greenstreet:2016}
Considering a 4 Gyr history for the scattering population, a transient stick from time $t$ must stick to a resonance for a timescale longer than (4 Gyr - $t$) to still be seen as resonant today. 
The similarity in the exponents for the population decay and the sticking probability means that the much larger scattering population at early times is essentially balanced by the decreased probability of the very long sticking timescales required for those objects to remain resonant until today. The result is that the distribution of currently observed transiently stuck resonant objects from a decaying scattering population is expected to have roughly flat distribution of $log(t_{stick})$ \citep{Yu:2018}. 

The expected transiently stuck resonant populations in Figure~\ref{f:extrapolation} are based on a 1 Gyr simulation and resonance sticking timescales from $10^5-10^9$ years. If we assume a relatively flat distribution of sticking timescales in $log(t)$ from the above considerations, extending the population estimates to include transient sticks from as long as 4 Gyr ago will increase the resonant population estimates by $\sim15\%$. 
This does bring the theoretically expected stuck 9:1 population into slightly better agreement with the observationally derived population estimate.
However, we do not expect `transiently' stuck resonant objects from early in the solar system's history to be a large fraction of the current resonant population unless one or more of the following is shown to be true: the primordial scattering population was more than the expected one to two orders of magnitude larger than the current scattering population, the primordial scattering orbital distribution was such that resonance sticking was significantly more likely than in the current population, or the resonances themselves were sticker in the past (perhaps due to a larger orbital eccentricity for Neptune as in, e.g., \citealt{Levison:2008}).

It is also possible that the observed 9:1 objects were emplaced onto stable resonant orbits during the era of giant planet migration. 
Given the previous lack of observed 9:1 objects, this has not been examined in the literature for this resonance.
\citet{Pike:2017} did find test particles in many high-$a$ Neptune resonances (including the 9:1) in an analysis of the end state of a Nice model scenario for giant planet migration; however, the analysis did not differentiate between transiently stuck and primordially captured resonant objects and did not provide a 9:1 population estimate.  
It is possible that some scattering objects that stick to the 9:1 while Neptune is still migrating will evolve into more stable phase space within the resonances.
However, \citet{Nesvorny:2016} find that slow migration of Neptune results in objects with larger perihelia (similar to our 9:1 objects) being preferentially dropped out of distant resonances as Neptune migrates; thus it is unclear whether sticking during migration could increase the 9:1 population.
If additional observations of the 9:1 population do not improve the currently marginal agreement between its total population and the resonance sticking hypothesis, the possibility of primordial capture should be examined in more detail.

\subsection{Comparison to other distant n:1 resonant populations}\label{s:rescomparison}

Beyond 50~au, the $n$:1 populations with the best constrained population estimates are the
3:1 and 5:1 resonances.  
\citet{Alexandersen:2016} 
estimates that the 3:1 contains at least $\sim$1200 objects with $H_r<8.66$,  using a similar approach to the survey simulation in Section~\ref{s:survsim} where the orbital model is based on the population having eccentricities and inclinations similar to the objects detected in the survey. 
\citet{Pike:2015}
derive a 5:1 population estimate of $\sim 2000$
objects with $H_g<8$ from three observed 5:1 objects.  Assuming a typical $g-r$ color ($g-r=0.5$; see, e.g. \citealt{Alexandersen:2016}) and $\alpha$=0.8, this corresponds
to $\sim$4$\times10^4$ 5:1 objects with $H_r<8.66$ (the 95\% confidence lower bound on this estimate is $\sim10^4$ objects); this would be the largest resonant population known \citep[as suggested by][]{Gladman:2012}.

While a 3:1 population $\simeq$9 times smaller than the 9:1 is consistent with the transient resonant population ratios in Figure~\ref{f:extrapolation}, 
the observations indicate that the 5:1 population is nominally equal to or larger than the 9:1 population.
Thus the observed 5:1 population appears to be inconsistent with transient sticking, despite  
the fact that the observed  5:1 resonators appear to be only transiently present in the resonance, escaping on $10^8-10^9$ year timescales \citep{Pike:2015}.
It is possible that the 5:1 population estimate \citep[based on detections in the CFEPS survey,][]{Petit:2011,Petit:2017} is only large by chance and thus a significant overestimate; this is supported by the fact that the much larger OSSOS project discovered only one additional 5:1 resonator \citep{Bannister:2018}.
Future constraints on these resonant populations will shed more light on whether they are consistent with the resonance sticking hypothesis.

\section{Summary and Conclusions}\label{s:sum}

We have explored the short- and long-term dynamical evolution of the first two  known 9:1 resonant TNOs, which were detected by OSSOS. 
Both o4h39 and o5m72 have median resonant stability timescales of $\sim1$Gyr (Section~\ref{ss:longterm}), which is consistent with expectations for resonance sticking timescales out of the scattering population. 
However, approximately $20\%$ of the clones of each of these two objects remain resonant on 4 Gyr timescales, so a primordial origin cannot be ruled out based on current stability.
Given the population estimate for the current active scattering population and a model of resonance sticking from this population, we expect there to be $\sim1-4.5\times10^3$ transiently stuck objects in the 9:1 with $H_r<8.66$ (Section~\ref{s:stickmodel}).
We estimate that this number should only increase by $\sim15\%$ if the decay in the expected decay in the scattering population over the last 4 Gyr is considered (Section~\ref{s:primordial}).
Using the OSSOS survey simulator (Section~\ref{s:survsim}), we find that our two 9:1 detections imply an intrinsic population of $\sim4-30\times10^3$ 9:1 objects with $H_r<8.66$ ($D \gtrsim 100$ km), which is marginally consistent with the transient sticking hypothesis. 
A comparison of the OSSOS 9:1 population estimate to observational constraints on the 5:1 and 3:1 resonances yields mixed results; the ratio of 9:1 to 3:1 resonators is consistent with the resonance sticking hypothesis, but the 5:1 is more populated than expected (Section~\ref{s:rescomparison}).

We conclude that the two 9:1 objects detected by OSSOS are marginally consistent with the hypothesis that the 9:1 is populated by transient resonance sticking from the scattering population. However, this hypothesis should be revisited as population estimates of Neptune's distant resonances are refined in the future, because a primordial 9:1 population would have interesting implications for the Kuiper belt's dynamical history.

\vspace{-5pt}

\acknowledgments
We thank an anonymous referee for helpful comments that improved this manuscript.
KV and RMC acknowledge support from NASA Solar System Workings grant NNX15AH59G. KV acknowledges additional support from NASA grant NNX14AG93G. SML gratefully acknowledges support from the NRC-Canada Plaskett Fellowship. MTB appreciates support from UK STFC grant ST/L000709/1. The Center for Exoplanets and Habitable Worlds is supported by the Pennsylvania State University, the Eberly College of Science, and the Pennsylvania Space Grant Consortium. 

The authors recognize and acknowledge the sacred nature of Maunakea, and appreciate the opportunity to observe from the mountain. 
This project could not have been a success without the dedicated staff of the Canada--France--Hawaii Telescope (CFHT). 
CFHT is operated by the National Research Council of Canada, the Institute National des Sciences de l'Universe of the Centre National de la Recherche Scientifique of France, and the University of Hawaii, 
with OSSOS receiving additional access due to contributions from the Institute of Astronomy and Astrophysics, Academia Sinica, Taiwan.
This work is based on observations obtained with MegaPrime/MegaCam, a joint project of CFHT and CEA/DAPNIA and on data produced and hosted at the Canadian Astronomy Data Centre and on the CANFAR VOSpace.

\software{Swift \citep{Levison1994},Matplotlib \citep{Matplotlib}}

\facility{CFHT (MegaPrime)}.

\bibliography{91refs}
\bibliographystyle{aasjournal}

\end{document}